\documentclass[prd,preprint,nofootinbib]{revtex4}
\usepackage{graphicx}
\usepackage{amssymb}

\begin{document}

\title{Limit on the Neutrino Mass from the WMAP Three Year Data}

\author{  Masataka Fukugita, Kazuhide Ichikawa, Masahiro Kawasaki}
\affiliation{
Institute for Cosmic Ray Research, University of Tokyo,
Kashiwa 277 8582, Japan}
\author{Ofer Lahav}
\affiliation{Department of Physics and Astronomy, University College London, Gower Street, 
London WC1E 6BT, United Kingdom}

\date{\today}

\vskip10mm
\begin{abstract}
We derive an upper limit on the neutrino mass from the WMAP three-year
data alone by employing a deterministic minimisation method based on a
grid search in multi-parameter space. Assuming the flat $\Lambda$CDM
model with power-law adiabatic perturbations, we find $\sum m_{\nu} < 2.0$ 
eV in agreement with the result of the WMAP team.  This
result, the limit being nearly the same as that from the WMAP
first-year data, means that the fundamental limit on the neutrino mass
obtainable from the cosmic microwave background alone is already nearly
met, as anticipated from the previous analysis. We also clarify the
role of the polarisation data in deriving the limit on the neutrino
mass.
\end{abstract}
\maketitle


In \cite{Ichikawa:2004zi} it was shown that a valid limit on the
neutrino mass can be derived from the cosmic microwave background
(CMB) experiment alone.
This contrasted to the then believed view
that such a limit is obtained only when the CMB data are combined with
those from large-scale clustering of galaxies 
\cite{Tegmark:2003ud,Spergel:2003cb}: 
the limit given
by Tegmark et al \cite{Tegmark:2003ud} derived from the first year  
data of the {\it
Wilkinson Microwave Anisotropy Probe} (WMAP-1) \cite{Spergel:2003cb}  
is $\sum m_\nu<11$
eV at a 95\% confidence. This implies that massive neutrinos alone
may constitute the entire dark matter of the Universe. This conclusion
was supported by the analysis of Elgar\o y and Lahav \cite 
{Elgaroy:2003yh}.
Our limit from WMAP-1
alone, $\sum m_\nu<2.0$ eV (95\% CL)
obtained from a deterministic grid search, contradicts with that
of \cite{Tegmark:2003ud}. We ascribed this discrepancy to the  
possibility that
Markov chain Monte Carlo used in \cite{Tegmark:2003ud} was not long  
enough to sample
the likelihood function in the presence of strong parameter degeneracy.

The argument given by \cite{Elgaroy:2003yh} is based on the well-known 
fact that massive
neutrinos diminishes the small-scale power by free-streaming and this
leads to a limit on the neutrino mass. In this argument the fact was
not taken into account that massive neutrinos affect 
the CMB perturbations not only
by the modification of the power spectrum but also in some 
not so obvious ways: characteristically, the acoustic length scale
and the relative
heights of peaks are modified.  The shift of the position of the
acoustic peaks and the height of the first peak alone can be absorbed  into
the change of cosmological parameters, but
when the heights of the second and third peaks relative to the first
are modified, the shifts cannot be absorbed into that of the
cosmological parameters, leading to a limit on the neutrino mass.
This means that a limit on neutrino mass
could be obtained from the observed CMB multipoles only if neutrinos are
nonrelativistic at the recombination epoch, i.e., $m_\nu>0.6$ eV.
Several numerical analyses made recently seem to support this conclusion
\cite{MacTavish:2005yk,Hannestad:2006zg,Lesgourgues:2006nd}.

At the data release of the WMAP team for their three year observations
(WMAP-3) \cite{Spergel:2006hy}, they gave a limit on the  
neutrino mass $\sum
m_\nu<2.0$ eV (at a 95\% confidence) from their CMB data alone, using
a Markov chain Monte Carlo analysis. This agrees with our conclusion
in \cite{Ichikawa:2004zi}, where it was shown that the limit from  
WMAP-1, $\sum
m_\nu<2.0$ eV, could hardly be improved even with the increasing
quality of the CMB data.  In this report we attempt to verify this
explicitly by repeating the previous analysis with the WMAP-3 data using
the deterministic grid search algorithm. Incidentally, we try to
elucidate the role of the polarisation data if they could tighten the
constraint on the neutrino mass.


\begin{figure}
\begin{center}
\includegraphics[width=15cm] {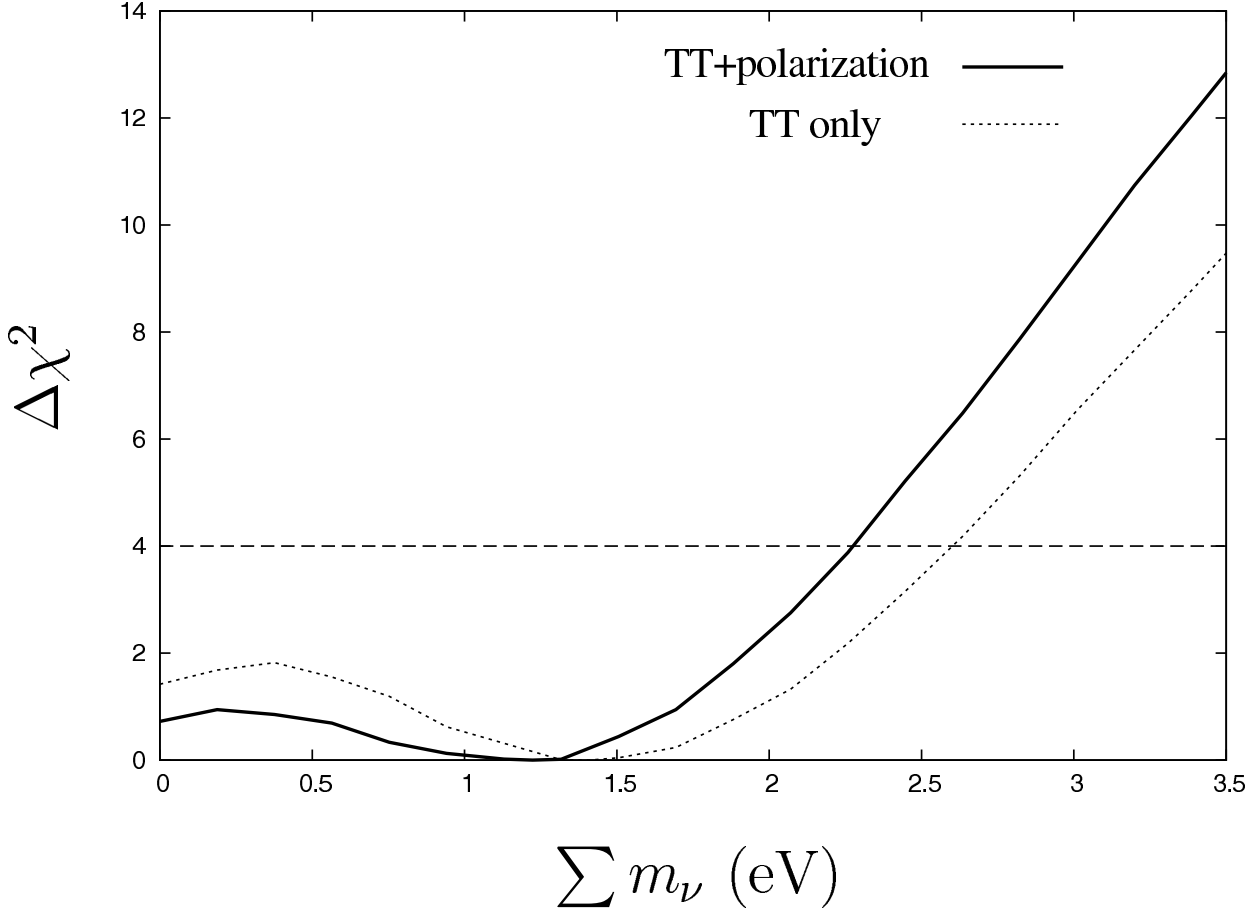}
\end{center}
\caption{The curve of $\chi^2(\omega_\nu)$ as a function
of the neutrino mass summed over generations. The solid
curve uses all available data from WMAP including
polarisations. The dotted curve is a constraint from only 
temperature-temperature correlations.} 
\label{fig:chi2}
\end{figure}


We take the flat $\Lambda$CDM model with adiabatic perturbations of
the power law spectrum. The model contains 6 cosmological parameters,
the mass density parameters ($\omega_i=\Omega_i h^2$) for cold dark
matter, baryonic matter and massive neutrinos, the normalised Hubble
constant $h$, the reionization optical depth $ \tau$, the scalar
spectral index of primordial mass perturbations $n_s$ and their
amplitude $A$. The neutrino mass density parameter $\omega_\nu$ is
related to neutrino masses by $\omega_ \nu = \sum m_\nu/$(94.1 eV) 
(the Fermi distribution is assumed) and
we assume three generations of massive neutrinos with a degenerate
mass.  Theoretical CMB power spectrum is calculated using the CMBFAST
code \cite{Seljak:1996is} and $\chi^2$ by the likelihood code of the
WMAP three-year data release 
\cite{Jarosik:2006ib,Hinshaw:2006ia,Page:2006hz}.
We calculate the $\chi^2$ function for a given 
$\omega_\nu$ by minimizing it over the 6 parameters. The
minimization is carried out with the grid search using the Brent
method \cite{brent} extended to multi-parameters, as described in
\cite{Ichikawa:2004zi}.

The result is shown in Fig.~\ref{fig:chi2}. The logarithm of the
likelihood  $-2\ln\cal{L}$ for the power spectra
including all TT, TE, EE and BB with the pixel-based method for low  
multipoles is shown by the solid line.
The $\chi^2_{\rm eff}$ for 3162 degrees of freedom at the vanishing  
neutrino mass is
1.037 for the parameters $\omega_b=0.0221$, $\omega_m=0.127$, $h=0.725$,
 $ \tau=0.091$ and $n_s=0.957$ in agreement with the WMAP-3  
analysis.

The limit corresponding to
$\Delta \chi^2 = 4$ (measured from the minimum) is $\omega_\nu \le
0.024$ or $\sum m_\nu \le 2.3$ eV. We obtain the upper limits at a
95\% confidence by an integration of the likelihood function ${\cal L}
= \exp \{-\Delta \chi^2/2 \}$,
\begin{eqnarray}
\omega_\nu \le 0.0215 \qquad {\rm or} \qquad \sum m_\nu \le
2.0 \ {\rm eV},
\label{eq:limit}
\end{eqnarray}
which corresponds to $m_\nu<0.67$ eV. A slight difference between
the limit from $\Delta \chi^2 = 4$ and likelihood integral is due to
the $\chi^2$ curve flatter than quadratic at small neutrino masses.
The actual minimum of $\chi^2$ occurs at a non-zero neutrino mass
$\sum m_\nu \approx 1.3$ eV, but $\chi^2$ relative to the vanishing
neutrino mass is less than one, meaning that the preference of
a finite neutrino mass is insignificant.

The limit we obtained agrees with that from the WMAP team, endorsing  
the validity of their
Markov chain Monte Carlo analysis. We may also compare
eq.~(\ref{eq:limit}) with that from WMAP-1 \cite{Ichikawa:2004zi}: $  
\omega_\nu \le 0.023$.
The limit changes little in spite of the significantly increased accuracy
of the CMB data, which agrees with the analysis given in \cite 
{Ichikawa:2004zi}.
This is close to the fundamental limit that could be obtained from
the CMB alone even if the data quality would be higher. A further
improvement of the limit needs external constraints, such as those 
from galaxy clustering.

\begin{figure}
\begin{center}
\includegraphics[width=12cm] {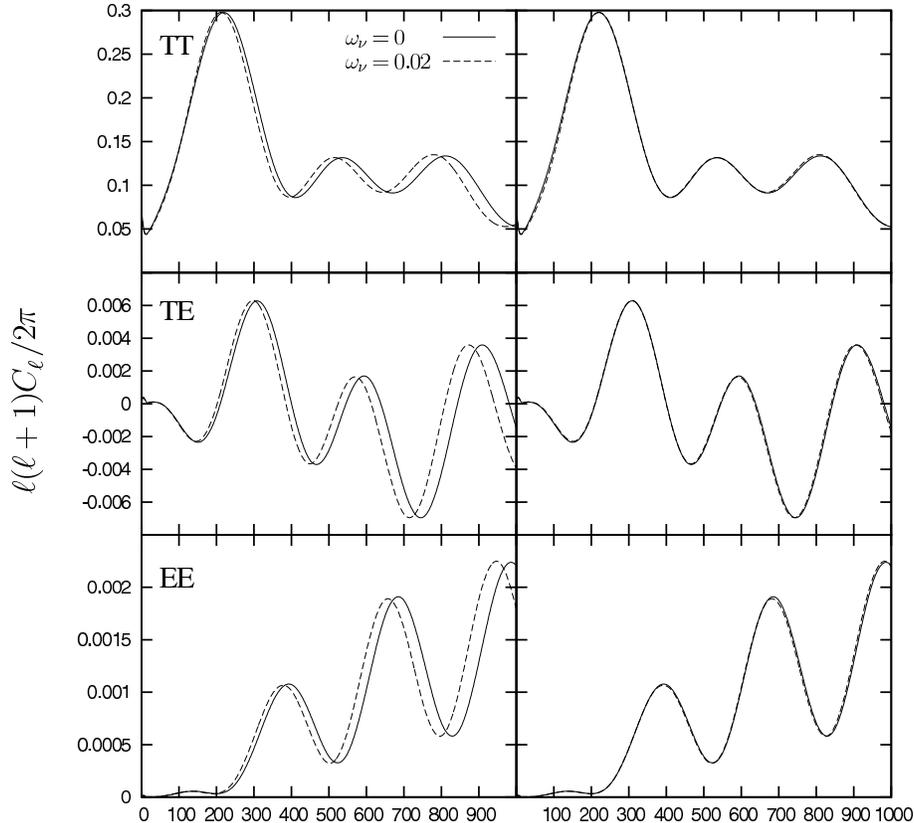}
\end{center}
\caption{TT, TE and EE correlators
for three cases: $\omega_\nu =$0 (the solid line) and  0.02 (the dashed
line) with the other parameters
fixed to the WMAP mean values. In the right panels
the abscissa is rescaled according to the
modification of the acoustic length that varies
with the neutrino mass.}
\label{fig:TE}
\end{figure}

We now briefly discuss the role of the polarisation data in deriving
the limit on the neutrino mass. We show in Figure 1 above the
$\chi^2$ curve for the TT correlation only (the dotted curve), which
gives $\omega_\nu \le 0.0250$ or $\sum m_\nu \le 2.4$ {\rm eV}, slightly
weaker than eq. (\ref{eq:limit}).  We
note, however, that there is no particular sensitivity in 
the TE or EE correlation to
the inclusion of the neutrino mass beyond that of the TT
correlation data.
The left column of the six panel figure, Figure \ref{fig:TE}, 
shows the TT, TE and EE correlators for $\omega_\nu=0$ and 0.02 with
the other cosmological parameters fixed. 
In the right-column panels we rescaled the abscissa in the way
that accounts for the modification of the acoustic length scale 
by a finite neutrino mass. The two curves almost coincide 
for all correlators, TT, TE and EE.  The WMAP data for TT are accurate
enough to marginally distinguish the two curves at approximately 
a 95\% confidence level. The data involving polarisation, however, 
do not reach this accuracy, so that they do not give
new information as to the neutrino mass. We do not expect
any effect of the neutrino mass that modifies polarisation
other than those through the scales in acoustic dynamics.  
The slight improvement on
the limit of the neutrino mass upon the inclusion of the polarisation
data, as we saw above, 
arises from the tightened constraint on $\tau$.  There is a
significant negative correlation between $\omega_\nu$ and $\tau$
(see  Figure 2d in \cite{Ichikawa:2004zi}). 
The absence of the polarisation data makes
$\tau$ more uncertain and rather
drives  
the $\chi^2$ minimum
to $\tau\approx 0$, which in turn pushes the neutrino mass towards a larger
value. In this way, the polarisation data serve to 
slightly improve the limit on the neutrino
mass through a constraint on the reionisation optical depth.

Our final remark is that the limit from the CMB data alone is the most
robust result in view of the 
uncertainties in the current large-scale
galaxy clustering data. They drive the best fit parameter of the
matter density either way depending upon whether one takes the SDSS 
or 2dFGRS data \cite{Spergel:2006hy}: the limit on the neutrino 
mass is also sensitive to
the choice of the galaxy clustering data between the two \cite{Fukugita06}
as well as how much
weight is given to these data.
To go beyond the present neutrino mass limit it would be imperative 
to combine external data, but this needs proper understanding 
of systematic errors involved in them.



\end{document}